\documentclass[11pt]{article}

\usepackage{graphicx,amssymb,amsfonts,amsmath,latexsym,natbib,color}
\usepackage[normalem]{ulem}

\newtheorem{theo}{Theorem}

\newcommand{\epr}{\hfill $\Box$\mbox{}\\ }

\newcommand{\bA}{\mathbf{A}}
\newcommand{\bB}{\mathbf{B}}
\newcommand{\bC}{\mathbf{C}}

\newcommand{\bI}{\mathbf{I}}
\newcommand{\bJ}{\mathbf{J}}

\newcommand{\bM}{\mathbf{M}}

\newcommand{\bR}{\mathbf{R}}
\newcommand{\bX}{\mathbf{X}}
\newcommand{\Un}{\mathbf{1}}
\newcommand{\Zero}{\mathbf{0}}

\newcommand{\bh}{\mathbf{h}}
\newcommand{\bm}{\mathbf{m}}
\newcommand{\bs}{\mathbf{s}}
\newcommand{\bu}{\mathbf{u}}
\newcommand{\bx}{\mathbf{x}}
\newcommand{\by}{\mathbf{y}}
\newcommand{\bz}{\mathbf{z}}

\newcommand{\bSigma}{\pmb{\Sigma}}
\newcommand{\bLambda}{\pmb{\Lambda}}
\newcommand{\blambda}{\pmb{\lambda}}
\newcommand{\bmu}{\pmb{\mu}}
\newcommand{\balpha}{\pmb{\alpha}}

\newcommand{\BS}{\mathbb{S}}
\newcommand{\BR}{\mathbb{R}}

\begin{document}

\title{Means and covariance functions for geostatistical compositional data: 
an axiomatic approach
\footnotetext{Authors are listed alphabetically. They contributed equally.}}

\date{}
\author{Denis Allard$^{a}$, Thierry Marchant$^{b}$\\
\small{$^{a}$Biostatistics and Spatial Processes, BioSP,  INRA, 84914, Avignon, France.}\\
\small{$^{b}$Ghent University, Ghent, Belgium.}
}

\maketitle

\noindent
Corresponding author:\\ 
D. Allard, BioSP, 84914 Avignon, FRANCE\\
Tel: +33\,432722171; Fax: +33\,432722182\\
{\tt denis.allard@inra.fr}
\vskip2.5mm

\begin{abstract}
This work focuses on the characterization of the central tendency of a sample of compositional data. It provides new results about theoretical properties of means and covariance functions for compositional data, with an axiomatic perspective. Original results that shed new light on the geostatistical modeling of compositional data are presented. As a first result, it is shown that the weighted  arithmetic mean is the only central tendency characteristic satisfying a small set of axioms, namely continuity, reflexivity and marginal stability. Moreover, this set of axioms also implies that the weights must be identical for all parts of the composition. This result has deep consequences on the spatial multivariate covariance modeling of compositional data.  In a geostatistical setting, it is shown as a second result that the proportional model of covariance functions (i.e., the product of a covariance matrix and a single correlation function) is the only model that provides identical kriging weights for all components of the compositional data. As a consequence of these two results, the proportional model of covariance function is the only covariance model compatible with reflexivity and marginal stability.
\end{abstract}

\bigskip

\newpage

\section{Introduction}
\label{sec:intro}

This work focuses on the characterization of the central tendency of a sample of compositional data and on 
consequences regarding its geostatistical modeling. Compositional vectors ``describe quantitatively the parts of some whole''
\citep{Egozcue2011}. They convey information about relative values of components, which are usually expressed in proportions or percentages. Their modeling and their analysis is therefore different from those of unconstrained multivariate vectors. In order to facilitate their analysis, it is natural to select a representative of the equivalent class \citep{Egozcue2011}. Therefore, compositional data with $p$  variables, $\bx=(x^1,\dots,x^p)$, has positive components that add up to a constant, say $\kappa$. Without loss of generality, one can set $\kappa=1$ for the rest of this paper.  Compositional data thus belongs
to the positive simplex of dimension $p-1$
\begin{equation}
  \label{eq:simplex}
\mathbb{S}^{p-1} = \{ (x^1,\dots,x^p)\}:\ x^k \geq 0\ 
\hbox{for}\  k=1,\dots,p,\ \hbox{and}\ x^1+\cdots+x^p = 1\}.  
\end{equation}

The central tendency of a sample of compositional data, which
is also a compositional data, is denoted by ${\bf M}(\bx_1,\dots,\bx_n)$.
\cite{Aitchison1989} states that the arithmetic mean is ``clearly useless
as a measure of location because it falls outside of the array of
compositions and is indeed very atypical of the data set'' and
that the normalized geometric mean (i.e., the back-transform of the arithmetic mean
of the log-ratios) ``serves equally well for curved data sets 
and for more linear and elliptical data sets''. 
On ternary sub-compositions of hongite \cite{Sharp2006} noted that  
in many instances the geometric mean also falls outside of the array of compositions, a fact already pointed out 
in \cite{Shurtz2000}. Probably the most striking example is the Hongite 2 artificial data sets where all 
samples are projected on a line parallel to one side of the triangle. The arithmetic
average is shown to belong to the same line, while the geometric average does not
\citep[Fig. 2]{Sharp2006}. 

\medskip

As an alternative to arithmetic or geometric means, \cite{Sharp2006} proposed the graph median.
It is built from the minimum spanning tree, which is the graph connecting all points
of the data set whose total length (sum of the length of the edges of the graph) 
is minimal. The distance used in the simplex is the half-taxi metric \citep{Miller2002}
$d(\bx,\bx') = 0.5\sum_{k=1}^p |x^k-x^{'k}|$. The graph median is then obtained by 
iteratively pruning the outermost branches of the tree until only one point
or a pair of points remain. This point or the mid-point between the pair of points 
is then the graph median. The minimum spanning tree is not always unique,
in which case a tie breaking rule is necessary. 
Very often, the graph median is one of the sample points of the data. Otherwise, it is the mid-point between 
two close data samples. By construction it is located in the innermost  region of the data-set, thus 
defining a central tendency \citep{Sharp2006}.
However, it is complicated to compute, and it cannot be easily related
to the estimation of a total quantity or indeed to most statistical 
or geostatistical analysis. Moreover, it is not continuous with respect to the 
data values. This alternative will not be considered in this work.

\medskip

The statistical analysis of compositional data has received great attention in the last three decades. 
Compositional data are often transformed into a new $p$-dimensional vector of log-ratios of the components
\citep{Aitchison1982, Aitchison1986,Pawlowsky2004,Egozcue2003}.  These transformations provide one-to-one mappings onto a 
real space, thereby allowing usual multivariate statistics methods to be applied to the transformed data. 
Any statement made in the transformed space is easily translated  back into the compositional space. 
\cite{Billheimer2001} and \cite{Pawlowsky2002} showed that the simplex $\mathbb{S}^{p-1}$ 
equipped with the scalar product 
\begin{equation}
  \label{eq:scalar_prod}
  \langle \bx,\by \rangle_A = \frac{1}{p}\sum_{i=1}^p \sum_{j=i+1}^p \ln \frac{x_i}{x_j} \ln \frac{y_i}{y_j} 
= \sum_{i=1}^p \ln \frac{x_i}{g(\bx)} \ln \frac{y_i}{g(\by)}, 
\quad  \bx,\by \in \mathbb{S}^{p-1},
\end{equation}
where $g(\bx) = [x^1x^2 \dots x^p]^{1/p}$ is the geometric mean of the components of $\bx$, 
induces a distance and  thus a geometry on the simplex, called the Aitchison geometry. 
An orthonormal basis on the simplex was proposed in 
\citep{Egozcue2003,Mateu2011}
where the data $(x^1,\dots,x^p)$  are transformed by means of the isometric log-ratio (ilr) transformations 
\begin{equation}
  \label{eq:ilr}
u^i = [i(i+1)]^{-1/2} \ln \frac{x^1 x^2 \dots x^i}{(x^{i+1})^i}, \quad i=1,\dots,p-1,  
\end{equation}
with $\bu = (u^1,\dots,u^{p-1}) \in \BR^{p-1}$. 
Some properties of compositional data and their parameters can be worked 
out directly within the Aitchison geometry on the simplex. However, when back-transforming 
these to the standard Euclidean space of the raw data, unexpected behavior may arise.
Following \cite[pp. 35--36]{Mateu2011}, let us consider the following example, with
$\bx = (0.6 ,0.3, 0.1)$ and $\bx' = (0.3, 0.3, 0.4)$ being two compositions in $\mathbb{S}^{2}$. Their ilr transforms
$\bu = (0.490, 1.180)$ and $\bu' = (0.000, -0.235)$ correspond to orthonormal coordinates with respect to the ilr transformation. 
We can then apply standard operations on these coordinates. For example, their arithmetic mean is
$\bar\bu = (0.245, 0.4725)$. Once back-transformed by the inverse operations of Eq. (\ref{eq:ilr}), we obtain
$$\hbox{ilr}^{-1}(\bar\bu) =  \tilde \bx = (0.459, 0.325, 0.216),$$
which is nothing but the normalized geometric mean of $\bx$ and $\bx'$.  
An unexpected and intriguing fact is that
even though the second coordinates are equal to 0.3 for both data, the second coordinate of $\tilde \bx$ is increased 
by 8.3\%.

\medskip

In the simplex, whenever one component is increased (resp. decreased), some or all other components
must decrease (resp. increase) in order to satisfy the sum constraint, a fact that has consequences on the conditions one 
wishes to impose on ${\bf M}$. Consider a dataset $\bx_{1}, \ldots, \bx_{n}$ of 
compositional data with $q$ variables (i.e., $\bx_{i} = (x_{i}^{1}, \ldots, x_{i}^{q})$)
to be grouped into $p<q$ variables in two different ways, thereby defining two new datasets $\by_{1}, \ldots, \by_{n}$ and $\bz_{1}, \ldots, \bz_{n}$. We further assume  that the first variable is identical in the two groupings, i.e., $y_{i}^{1}=z_{i}^{1}$, $i = 1, \ldots, n$. For instance, with $q=4$ and $p=3$, one grouping is $1, \{2,3\}, 4$ and the other one is $1, 2, \{3,4\}$. One does not expect the mean of the first variable to depend on 
the grouping  of the other variables. This is the case for the arithmetic mean. Indeed,
for the first grouping, the $k$th component of the arithmetic mean is
$M_{a}^k(\by_{1}, \ldots, \by_{n}) = n^{-1}\sum_{i=1}^n y^k_i$,
and a similar expression holds for  $M_{a}^k(\bz_{1}, \ldots, \bz_{n})$.
Since the first variable is common to both datasets, it is clear that $M_{a}^1(\by_{1}, \ldots, \by_{n})=M_{a}^1(\bz_{1}, \ldots, \bz_{n})$.
The arithmetic mean is said to satisfy marginal stability.

Let us now compute the normalized geometric mean independently for each variable. Then, for the first grouping
$$M_{g}^k(\by_{1}, \ldots, \by_{n}) = \frac{\left( \prod_{i=1}^n y^k_i\right)^{1/n}}{\sum_{l=1}^p
\left( \prod_{i=1}^n y^l_i\right)^{1/n}},$$
for $ k=1,\dots,p$, with a similar expression for the second grouping. 
Since all variables are involved in the computation of each component of the normalized
geometric mean, equality of the mean of the first variable for the two groupings will be different, 
i.e.,  $M_{g}^1(\by_{1}, \ldots, \by_{n}) \neq M_{g}^1(\bz_{1}, \ldots, \bz_{n}).$
In other words, the normalized geometric mean does not  satisfy marginal stability. A similar result is obtained for any
log-ratio transform. Marginal stability is an important property that
one may choose to impose or not when analyzing compositional data. It plays a critical role in our results.

\medskip

For applications, such as oil and mining industry or soil remediation, where compositions correspond 
to fractions of soil type or rock type, the above examples are puzzling. Based on samples of compositional samples,
practitioners working in these fields need to compute reliable estimates of the total amount of material for a given volume, or equivalently, estimates of the mean of the fraction. Unanswered questions arise: Under which conditions is marginal stability always  satisfied on central tendencies?  In a geostatistical setting, what are the conditions for having unbiased estimates of means and what are the multivariate covariance functions compatible with marginal stability?

\medskip

Through an axiomatic perspective, we provide answers to these questions. New mathematical results about theoretical properties of central tendencies and compatible covariance functions for compositional data are presented. These results shed new light on its  geostatistical modeling. We start from a set of axioms that correspond to  quite natural conditions: continuity, reflexivity (which will be defined later) 
and marginal stability.  Other sets of axioms would lead to 
different mathematical properties. For example, subcompositional dominance is violated by the Euclidean distance between compositional data
\citep{Egozcue2011}. It is not the purpose of the present work to discuss the relevance of sets of axioms for the analysis of data. 
We merely derive ``if and only if'' relationships between one given set of axioms and some mathematical properties, 
which we believe are useful in some situations. 

\medskip

As a first result, it will be shown that the weighted  arithmetic mean is the only central tendency characteristic satisfying reflexivity and marginal stability. Moreover, the weights must be identical for all components of the compositional vector.
This first result  holds in the standard Euclidean space of the raw data. It has deep consequences on the multivariate covariance modeling. 
It is well known (and easy to verify) that, if the multivariate covariance function belongs to the family of 
proportional models (i.e., it is the product of a covariance matrix and a single correlation function), 
the kriging weights are identical for all variables \citep{Helterbrand1994}. It will be shown that the converse also holds for all multivariate random fields, compositional or not. As a second result on compositional data, it is then established that the proportional model of covariance functions is,
in the Euclidean space of raw data, the only covariance model satisfying reflexivity and marginal stability.  This second result
does not necessarily hold within the Aitchison geometry on the simplex.
\medskip

The paper is structured as follows. Section~\ref{sec:axioms_multi} provides a quick presentation of the axiomatic characterization of the possible
definitions of means. Section \ref{sec:axioms_comp} presents our first main result on the mean of compositional data. 
In Sect. \ref{sec:cov}, the consequences of this theorem on the covariance functions are shown. These results are then discussed in Sect. \ref{sec:discussion}.

\section{A primer  to axiomatic definitions of  means}
\label{sec:axioms_multi}

\subsection{Means for univariate data}

Many ways exist of summarizing a sample $x_1,\dots,x_n$ of a variable defined on an interval $E \subseteq \BR$ 
into a single value, usually called the mean, and sometimes also called central tendency. 
The most widely used are the arithmetic, geometric or harmonic means, the mode,
the median, and more generally  any desired quantile. They all satisfy different properties. 
For example, the arithmetic mean $\bar{x} = n^{-1}\sum_{i=1}^n x_i$ is the real number $m$ minimizing $\sum_{i=1}^n (m-x_i)^2$. 
It is unique, easy to compute, but sensitive to  large values or outliers.
The median is the real number  $m$ minimizing $\sum_{i=1}^n |m-x_i|$. It is sometimes not unique, 
but robust to large values and outliers.  An axiomatic characterization of the arithmetic mean was 
obtained as early as \cite{Kolmogorov1930}.
Let us denote by $M(x_1,\dots,x_n)$ a mapping from $E^n \to E$, which will be the central
tendency, or the mean, of the sample, and let us impose the following quite natural conditions, called axioms.
Kolmogorov's characterization follows.
\begin{enumerate}
\item[K1]  {\em Continuity and strict monotonicity:}
$M(x_1,\dots,x_n)$ is continuous and strictly monotonic in each of its arguments. 
Strict monotonicity states that if one of the values $x_{i}$ increases, the central tendency must also increase. 
Continuity imposes that small variations in one of the values $x_{i}$ leads to small variations of $M$.
\item[K2]  {\em Symmetry:} $M(x_1,\dots,x_n)$ is symmetric, i.e., it is identical for any permutation of the sample.
\item[K3]  {\em Reflexivity:} The central tendency of identical values is equal to 
their common value, i.e.,
$M(x,\dots,x)=x$.
\item[K4] {\em Associativity:}
A subset of the sample can be replaced by its central tendency with no effect
on the total central tendency:
$$M(x_1,\dots,x_m,x_{m+1},\dots,x_n) = M(x_*,\dots,x_*,x_{m+1},\dots,x_n)$$
where $x_* = M(x_1,\dots,x_m)$.
\end{enumerate}

\begin{theo}[Kolmogorov, 1930]
  Conditions K1 to K4 hold if and only if the central tendency $M$ has the form
  \begin{equation}
    \label{eq:Kolmogoroff}
    M(x_1,\dots,x_n) = \phi^{-1}\left( \frac{\phi(x_1)+\cdots+\phi(x_n)}{n} \right),
  \end{equation}
where $\phi$ is a continuous strictly increasing function on $E$, called the generating function.
\end{theo}

Functions $M$ having the form of Eq.~(\ref{eq:Kolmogoroff}) are 
called quasi-arithmetic means in the functional equation literature
\citep{Aczel1989, Matkowski2010}. It is worth noting that neither the mode nor the median belong to this family.
The median is not continuous,  while neither the mode nor the median satisfy K4.
Quasi-arithmetic means cover a wide range of well known means:
if $\phi(x)=x$, $M$ is the arithmetic mean. On $E=(0,+\infty)$, $\phi(x)=\ln x$
leads to the geometric mean, while $\phi(x)=x^{-1}$ leads to the harmonic mean.
More generally,  when $\phi(x)=x^{\alpha}, \ \alpha \neq 0$ and  $x \in (0,+\infty)$ the associated mean is called
the power mean.

\subsection{Means for multivariate data}

For multivariate data, each data is a vector of length $p$, i.e., $\bx_i = (x_i^1,\dots,x_i^p)$, with $i=1,\dots,n$.
In this case, the central tendency is a vector $\bM=(M^1,\dots,M^p)$, where $M^k$ is a mapping from $E^{n} \to E$. 
Possibilities are numerous.  A tempting simplification is to impose that each component $M^k$
depends only  on the values of the corresponding variable.  In this case one builds one central tendency per variable 
(e.g., the arithmetic mean or the median), independently of all other variables.
When applying Kolmogorov's Theorem to each variable independently, the associated central tendency
$M^k$  is a quasi-arithmetic mean if and only if conditions K1-K4 hold. When $E \subseteq \BR$,
the multivariate arithmetic mean is the point $\bM(\bx_1,\dots,\bx_n) = \bm  \in \BR^p$ 
minimizing  $\sum_{i=1}^n ||\bm-\bx_i||_p^2$, where $||\cdot||_p$ is the Euclidean distance in $\BR^p$. 
In this case, each component of $\bM(\bx_1,\dots,\bx_n)$ is the arithmetic mean of the corresponding variable.

\medskip

Although being natural and probably widely used, the componentwise simplification is by no means the only mathematical possibility.  
For instance, in analogy to  the definition of the median for univariate data, the multivariate median is
the $p$-dimensional vector $\bm$ minimizing $\sum_{i=1}^n ||\bm-\bx_i||_p$. It is 
unique whenever the points are not colinear \citep{VardiZhang2000}.
Its $k$-th coordinate, $m^k$, depends on the values of all coordinates of the samples $\bx_1,\dots,\bx_n$.
Many alternative multivariate medians can be defined, such as those based on the notion of statistical depth \citep{Liu1999,Zuo2000} or
the graph median proposed in \cite{Sharp2006}.

\section{An axiomatic characterization of the mean for compositional data}
\label{sec:axioms_comp}

As seen in the Introduction, there exist many possible approaches to defining a mean for compositional data, 
and they all satisfy different properties.  Inspired by the axiomatic approach briefly summarized in Sect. 
\ref{sec:axioms_multi}, a new characterization theorem for compositional data is presented. It relies on 
functional equation arguments and does not necessitate the use of inner products and distances. It is independent of
any geometry on the simplex.

\medskip

Consider  a sample of fixed size $n$ of compositional data
belonging to the positive simplex $\mathbb{S}^{p-1}$, 
corresponding to a regionalized variable $\bx(\cdot)$ defined in
$\BR^d$, sampled at sites 
$\bs_1,\dots,\bs_n$. For the sake of simplicity, 
we will write $\bx_i=\bx(\bs_i)$. We want to characterize the mapping $\bM=(M^{1}, \ldots, M^{p})$,
which associates to 
each data-set  $(\bx_{1}, \ldots, \bx_{n})$ a $p$-dimensional mean vector 
$\bM(\bx_{1}, \ldots, \bx_{n})$ with 
components $M^{k}(\bx_{1}, \ldots, \bx_{n})$, for  $ k=1,\dots,p$.
Notice that at this stage $M^{k}(\bx_{1}, \ldots, \bx_{n})$ can depend 
on $\bx_{1}, \ldots, \bx_{n}$  and not just on $x_{1}^{k}, \ldots, x_{n}^{k}$, so that 
the definition of $\bM$ is very general. 

We will only consider  mappings $\bM$ such  that $\sum_{k=1}^{p} M^{k}(\bx_{1}, \ldots, \bx_{n}) = 1$. In order to account for the fact that Kriging can result in negative weights and that this, in turn, can yield negative estimates for some specific data sets, we will not assume from the onset that $ 0 \leq M^{k}(\bx_{1}, \ldots, \bx_{n}) \leq 1$. The set of all real vectors $\bx$ of size $p$ satisfying $\sum_{k=1}^{p} x^k=1$ will be denoted by ${\cal S}^{p-1}$. It is a superset of the simplex $\BS^{p-1}$.

Conditions imposed for the characterization of the central tendency of 
spatial compositional data differ from those imposed for Kolmogorov's 
characterization. Among Kolmogorov's conditions, we retain only two: reflexivity and continuity.
As discussed in the previous section, marginal stability is also imposed. 
Our main theorem, characterizing the means for compositional data, follows.

\begin{enumerate}

\item[C1] Reflexivity: 
for all $\bx \in \BS^{p-1}$, $\bM(\underbrace{\bx,\dots,\bx}_{n}) =\bx$.

\item[C2] Marginal stability: for any $k=1,\dots,p$, any $i=1,\dots,n$
and any $\bx_{1},\dots,\bx_{n}$, $\bx'_{i}$ in $\BS^{p-1}$, if $x_{i}^k=x_{i}^{'k}$, then
$$M^k(\bx_{1},\dots,\bx_{i},\dots,\bx_{n}) = M^k(\bx_{1},\dots,\bx'_{i},\dots,\bx_{n}).$$

\item[C3] Continuity: $\bM$ is continuous in each argument.

\end{enumerate}

In presence of spatial correlation, spatial symmetry is in general not desired, but can be an option. 
For the sake of completeness it is recalled. 
\begin{enumerate}

\item[C4] Symmetry: 
$$M^k(\bx_{1},\dots,\bx_{n}) = M^k(\bx_{\sigma(1)},\dots,\bx_{\sigma(n)})$$
for any permutation $\sigma$ of $\{1,\dots,n\}$ and any $\bx_{1},\dots,\bx_{n}$ of $\BS^{p-1}$. 

\end{enumerate}

We are now ready to state our first result. In part A, we do not impose that the components of $\bM(\bx_{1}, \ldots, \bx_{n})$ be nonnegative (i.e. $\bM: (\BS^{p-1})^n \to {\cal S}^{p-1}$). In part B, this nonnegativity constraint is imposed (i.e. $\bM: (\BS^{p-1})^n \to \BS^{p-1}$).

 \begin{theo}[Characterization of the mean]
    \label{theo:maintheo}
Consider compositional data $(\bx_{1}, \ldots, \bx_{n})$ with  $p \geq 3$ and $n \geq 2$. 
\begin{enumerate}
	\item[A)]  The mapping $\bM: (\BS^{p-1})^n \to {\cal S}^{p-1}$ satisfies conditions C1-C3 if and only if, 
	for $k \in \{1, \ldots, p\}$, 
	\begin{equation}
	\label{eq:Mwav}
	M^{k}(\bx_{1}, \ldots, \bx_{n,}) = \sum_{i=1}^{n} \lambda_{i} x_{i}^{k},
	\end{equation}
	for some real numbers $(\lambda_{1}, \ldots, \lambda_{n}$)  satisfying 
	$\sum_{i=1}^n \lambda_i =  1.$
	\item[B)] The mapping $\bM: (\BS^{p-1})^n \to \BS^{p-1}$ satisfies conditions C1-C2 if and only if, 
	for $k \in \{1, \ldots, p\}$, 
	\begin{equation}
	\label{eq:Mwav2}
	M^{k}(\bx_{1}, \ldots, \bx_{n,}) = \sum_{i=1}^{n} \lambda_{i} x_{i}^{k},
	\end{equation}
	for some non-negative real numbers $(\lambda_{1}, \ldots, \lambda_{n}$)  satisfying 
	$\sum_{i=1}^n \lambda_i =  1.$  
\end{enumerate}
Furthermore, if  symmetry holds, then $\lambda_{i} = 1/n$ for all $i \in \{1, \ldots, n\}$ and $\bM(\bx_{1}, \ldots, \bx_{n}) \in \BS^{p-1}$.
\end{theo}
The key ingredients of the proof, presented in the Appendix, are marginal stability and sum to 1.
Marginal stability imposes that, for each $k=1,\dots,p$, $M^k$ 
depends only on $x_1^k,\dots,x_n^k$. Then, the sum to 1 condition implies that the generating function $\phi$ must be the identity function, i.e., that the only means are linear combinations. Clearly, the same result holds if the data sum to another constant than 1. Notice that when the mapping $\bM$ is bounded, continuity of the mapping (axiom C3) is not necessary for the proof. However, since linear combinations are continuous, continuity is satisfied in both cases. As a consequence, the only means satisfying C1-C3 are linear combinations, whether or not $M^{k}(\bx_{1}, \ldots, \bx_{n,})$ is forced to belong to $\BS^{p-1}$.

\medskip

Theorem \ref{theo:maintheo} holds whenever there are at least three components.
With two components there is only one independent variable, and hence
many more means satisfying axioms C1-C3 exist. As an example, it can easily be shown 
that the Kolmogorov means characterized in  Eq. (\ref{eq:Kolmogoroff})
 $$M^{k}(\bx_{1}, \ldots, \bx_{n}) = \phi^{-1} \left( \sum_{i=1}^{n} \lambda_{i} \phi(x_{i}^{k}) \right)$$
 also satisfy marginal stability and sum to 1 whenever $\phi : \ [0,1]
\rightarrow \mathbb{R}$  is such that $\phi(t)=1-\phi(1-t)$, with $\phi(0)=0$ and $\phi(1)=1$. Indeed,
\begin{eqnarray*}
M^{1}(\bx_{1}, \ldots, \bx_{n}) + M^{2}(\bx_{1}, \ldots, \bx_{n}) & =&
   \phi^{-1} \left( \sum_{i=1}^{n} \lambda_{i} \phi(x_{i}^{1}) \right) + \phi^{-1} \left( \sum_{i=1}^{n} \lambda_{i} \phi(1-x_{i}^{1}) \right) \\
 & = &\phi^{-1} \left( \sum_{i=1}^{n} \lambda_{i} \phi(x_{i}^{1}) \right) + \phi^{-1} \left( \sum_{i=1}^{n} \lambda_{i} (1-\phi(x_{i}^{1})) \right)\\ 
  & = &\phi^{-1} \left( \sum_{i=1}^{n} \lambda_{i} \phi(x_{i}^{1}) \right) + \phi^{-1} \left( 1 - \sum_{i=1}^{n} \lambda_{i} \phi(x_{i}^{1}) \right) \\
\end{eqnarray*}
Since $\phi(t)=1-\phi(1-t)$ implies $\phi^{-1}(u)=1-\phi^{-1}(1-u)$ for $0 \leq u \leq 1$, the last equality implies
$$ M^{1}(\bx_{1}, \ldots, \bx_{n}) + M^{2}(\bx_{1}, \ldots, \bx_{n}) = \phi^{-1} \left( \sum_{i=1}^{n} \lambda_{i} \phi(x_{i}^{1}) \right) + 
1- \phi^{-1} \left(  \sum_{i=1}^{n} \lambda_{i} \phi(x_{i}^{1}) \right) =1.$$

\medskip

The case $p=2$ is not of central interest for the analysis of compositional data. In the rest of this work, it is thus assumed that 
$p \geq 3$.

\section{Geostatistical compositional data}
\label{sec:cov}

\subsection{Characterization of compatible covariance models}

Within a (geo-) statistical setting, the condition $\sum_{i=1}^n \lambda_i = 1$ in Eq. (\ref{eq:Mwav}) corresponds to an unbiasedness condition. 
Theorem  \ref{theo:maintheo} states that  means of compositional data satisfying axioms C1-C3 correspond to linear unbiased estimators of the population mean.
However, this theorem does not provide any criterion for choosing the weights in Eq.~(\ref{eq:Mwav}). In particular, it does not make explicit reference to the location of the data.

\medskip

In geostatistics, it is well known that, provided the covariance 
function is known, the Best Linear Unbiased Estimator of the  
mean vector is the generalized-least-squares (gls) estimator \citep{Cressie1993}. It is also referred to as ``kriging of the mean'' in \cite{Wackernagel} and \cite{Chiles2012}, a term that will be used in the rest of this work. In the following, ``kriging weights'' is a (slight) abuse of language to refer to the weight of the gls estimator of the mean.  The fact that the kriging weights when estimating the mean must be identical for all variables under axioms C1-C3 has deep consequences on the covariance modeling of spatial compositional data, which are 
now examined in detail. The usual geostatistical setting is considered. The compositional data $(\bx_{1}, \ldots, \bx_{n,})$ 
is a sample of a second order stationary multivariate random field ${\cal X}(\bs)$ with $\bs \in \BR^d$. For the simplicity of exposition, we
consider $p \geq 3$ and $n \geq 2$. 
Under the assumption of second order stationarity,
the multivariate covariance model of ${\cal X}(\bs)$ is defined by a matrix of functions
\begin{equation}
\left( \begin{array}{ccc} C_{11}(\bh) & \cdots & C_{1p}(\bh) \\  
\vdots & \ddots & \vdots \\
C_{p1}(\bh) & \cdots & C_{pp}(\bh) \end{array}\right),\ \bh\  \in \BR^d,
\end{equation}
which must be positive definite \citep{Chiles2012}.
For a given set of $n$ locations $(\bs_1,\dots,\bs_n)$, this model induces a  $np \times np$ multivariate covariance  block-matrix of the form
\begin{equation}
\bC = \left( \begin{array}{ccc} \bC_{11} & \cdots & \bC_{1p} \\  
\vdots & \ddots & \vdots \\
\bC_{p1} & \cdots  & \bC_{pp} \end{array}\right),
\end{equation}
where each  matrix $\bC_{kl}$  is such that its elements are $[\bC_{kl}]_{ij} = C_{kl}(\bs_j -  \bs_i)$, with $1 \leq k,l \leq p$ and $1 \leq i,j \leq n$. 

\medskip

Kriging weights depend on the multivariate covariance model and on sample locations. For some specific combinations of covariance models and sample locations, some of the kriging weights can be negative. 
%In this case, there is therefore a possibility that $M^k(x^k_1,\dots,x^k_n) \not\in [0,1]$ for some component $k$ and some very specific data set. 
Obviously, for all practical purposes, a basic requirement is that the mean $\bM(\bx_{1}, \ldots, \bx_{n})$ lies in the simplex $\BS^{p-1}$.
%Since the sum of the components of $\bM$ is equal to 1, requiring $\bM(\bx_{1}, \ldots, \bx_{n}) \in \BS^{p-1}$ is equivalent to require that $M^k (x^k_{1}, \ldots, x^k_{n})\geq 0$ for $k=1,\dots,p$. Considering that compositional data are positive,  the condition $M^k (x^k_{1}, \ldots, x^k_{n}) \geq 0$ will almost always be verified. Indeed, 
Since $M^k(x^k_{1}, \ldots, x^k_{n})$ is the central tendency of composition $k$, it is  expected that it lies within the interval defined by the minimum and maximum of the observed values of $\bx^k$, which is  a sufficient condition for $0 \leq M^k(x^k_{1}, \ldots, x^k_{n}) \leq 1$.   However, in presence of negative weights it is mathematically not impossible that for some very specific datasets 
one gets $M^k(x^k_{1}, \ldots, x^k_{n}) \not \in [0,1]$, which is not acceptable in our context. In Section \ref{sec:restricted}, we will discuss how the condition  $\bM(\bx_{1}, \ldots, \bx_{n}) \in \BS^{p-1}$ can be enforced. At this point, let us simply notice that imposing  nonnegative weights is a sufficient condition.

\medskip

It is well known (and easy to verify) that if the multivariate covariance function belongs to the family of proportional models (i.e., the product of a covariance matrix and a single correlation function), the kriging weights are identical for all variables \citep{Helterbrand1994}. The following Theorem shows that the converse also holds. It is first stated independently of the compositional data setting considered so far. A formal link to the compositional data setting will be made later.
\begin{theo}
\label{theo:maintheo2}
Let us consider a second-order stationarity multivariate random field of dimension $p \geq 2$. 
\begin{enumerate}
\item[A)] For all $n \geq 2$ and all  $(\bs_1,\dots,\bs_n)$, the co-kriging weights of the $p$ means are equal
for all variables if and 
only if the multivariate covariance model is a proportional model, i.e., for $1 \leq k,l \leq p$
$$C_{kl}(\bh) = \sigma_{kl} \rho(\bh),\ \bh \in \BR^d,$$
for some  $p \times p$ covariance matrix $\bSigma = [\sigma_{kl}]_{1}^{p}$ and some correlation function $\rho(\bh)$. 
\item[B)] Moreover, the same result holds if nonnegativity of the kriging weights is imposed.
\end{enumerate}

\end{theo}
The proof of this Theorem is deferred to the Appendix. Particular cases of this model are absence of correlation between variables, with $\sigma_{kl}=0$ for all $k \neq l$. This particular case must be ruled out when considering compositional data since negative correlation between some variables is implied by the fact that the parts sum to one. 

\medskip

Theorems \ref{theo:maintheo} and
\ref{theo:maintheo2} considered together 
imply that, when estimating the central tendency of geostatistical compositional data,  
the only minimum variance estimator satisfying reflexivity, marginal stability and continuity is the kriging of the mean with identical weights for all compositions, whether or not imposing nonnegative weights to enforce $\bM(\bx_1,\dots,\bx_n)$ to be in the simplex is needed. Moreover, the only multivariate covariance function model satisfying these requirements is the proportional model
$$\bC(\bh) = \bSigma \otimes \rho(\bh)$$
where $\rho(\bh)$ is a correlation function and $\bSigma$ a valid covariance matrix for compositional data.  This is now formally stated in the following Theorem.

\begin{theo}
Let ${\cal X}(\bs)$  be a multivariate compositional random field as described above and
let $\bM  = (M^1,\dots,M^p)$, with $\bM(\bx_1,\dots,\bx_n) \in \BS^{p-1}$, be a compositional mean such that
$\hbox{\rm tr}\{ \hbox{\rm Var}(\bM) \}$ is minimal over $\BS^{p-1}$. Then, 
the following propositions are equivalent:
\begin{enumerate}
\item $\bM$   satisfies C1-C3.
\item For each variable $k=1,\dots,p$, $M^k(\bx_1,\dots, \bx_n)$ is a linear combination of $(x_1^k,\dots,x_n^k)$, i.e.,
$M^k = \sum_{i=1}^n \lambda_i x_i^k$,  where the weights are the same for all variables. 
The $n$-dimensional vector of weights $\blambda = (\lambda_1,\dots,\lambda_n)^\top$ is the solution of a unique (possibly constrained) kriging system.
\item The multivariate covariance function model is a proportional model.
\end{enumerate}
\label{theo:theo4}
\end{theo}

\noindent {\bf Proof}. The proof consists in collecting the results established in Theorems \ref{theo:maintheo} and \ref{theo:maintheo2}.
\begin{enumerate}
\item[a)] According to Theorem \ref{theo:maintheo}, ``$\bM$ satisfies C1-C3'' is equivalent to ``$M^k(\bx_1,\dots, \bx_n)$ is a linear combination of 
$(x_i^k,\dots,x_n^k)$ only. Moreover, the weights are identical for all variables $k=1,\dots,p$''. Therefore, imposing 
$\hbox{\rm tr}\{ \hbox{\rm Var}(\bM) \}$ to be minimal on $\BS^{p-1}$ is equivalent to imposing ${\rm Var}(M^k)$ to be minimal for each $k=1,\dots,p$. Hence, 
it is equivalent to impose $M^k(\bx_1,\dots,\bx_n)$ to be equal to the (possibly constrained) kriging of the mean of  $(x_1^k,\dots,x_n^k)$, for some unique kriging system. 
Hence $1 \Leftrightarrow 2$.
\item[b)] In Theorem  \ref{theo:maintheo2} it is proven that $2 \Leftrightarrow 3$.
\end{enumerate}
In conclusion, the three statements are equivalent. \epr

\subsection{Enforcing the mean to be in the simplex}
\label{sec:restricted}

As already pointed out earlier, in practice the mean $\bM(\bx_{1}, \ldots, \bx_{n})$ obtained by unrestricted kriging of the mean (i.e., without imposing nonnegative weights) will almost always lie in the simplex. Indeed, since the correlation model $\rho(\bh)$ has been fitted on the dataset, it is expected that even though some of the weights are negative, these weights will be of small magnitude. Moreover, the combination of a high magnitude negative weight with extreme compositional value, leading to  $ M^k(x^k_{1}, \ldots, x^k_{n}) \not \in [0,1]$, is very unlikely.
In this case, part A) of Theorem \ref{theo:maintheo} and Theorem \ref{theo:theo4} apply. In the unlikely event where one gets   $ M^k(x^k_{1}, \ldots, x^k_{n}) \not \in [0,1]$ for some composition and some specific dataset, several possibilities are now discussed.

\begin{enumerate}
	\item  In order to force $\bM(\bx_{1}, \ldots, \bx_{n})$ to belong to $\BS^{p-1}$, \citet{Walvoort2001} proposed compositional kriging, for which the sum of the $p$ prediction variances is minimized, subject to unbiasedness, nonnegativity and sum to 1 of the estimated parts. Unfortunately, this approach leads to weights that i) depend on the observed values $\bX$ and ii) are different for each  part $k=1,\dots,p$. The compositional kriging proposed in \citet{Walvoort2001} is thus incompatible with axiom C2. 
	
	\item A sufficient condition is to impose that the weights $\lambda_1,\dots,\lambda_n$ are nonnegative. In this case, Part B) of Theorem \ref{theo:maintheo}  and Theorem \ref{theo:theo4} apply. From a geostatistical point of view, the kriging variance will be higher than that of the unrestricted kriging, but the condition $\bM \in \BS^{p-1}$ will be satisfied. 
	
	\item However, in our opinion, obtaining $M^k(x^k_{1}, \ldots, x^k_{n}) \not \in [0,1]$ for some composition $k$ can be interpreted as a lack of fit between the model and that specific dataset. Finding a more adequate model for the correlation function  $\rho(\bh)$ will fix the problem. There are several possibilities: decreasing the regularity of $\rho(\bh)$ at the origin (e.g. from quadratic to linear behavior),  decreasing its range, adding a small nugget effect.  Of course, any combination of these is also possible. 
\end{enumerate}

\section{Discussion}
\label{sec:discussion}

In this work several original results are established:
\begin{itemize}
\item Firstly, it is shown that for compositional data the only means that simultaneously satisfy  reflexivity, continuity and marginal stability
are weighted arithmetic means, to which kriging belongs. The simultaneous requirement of marginal stability 
and the sum to one is the main reason leading to this result. 
\item Secondly, a very general result in multivariate geostatistics is established. 
It is shown that the only multivariate covariance model 
for which the kriging weights are identical for all components is the proportional model. To the best of our knowledge, 
this result has not been shown earlier. It has consequences for the modeling of compositional data, but it also is of interest on its own. It is worth recalling here that \citet{Bogaert2002} showed that, for a categorical random field, the only valid multivariate covariance model for indicator co-kriging is also the proportional model. Any other form of the linear model of corregionalization is not permissible. 

\item Thirdly, the combination of these two results is that, within a geostatistical setting in the Euclidean 
space of the raw data, the only covariance model leading to kriging that 
satisfies simultaneously reflexivity and marginal stability is the proportional model. 
\end{itemize}
In summary, when performing the statistical analysis of compositional data in the standard Euclidean geometry of real space
it is impossible, at the same time, to satisfy axioms C1-C3 and to consider  complex modeling using log-ratio transformations, complex covariance models, or both.  It is a kind of impossibility result, in the spirit of Arrow's impossibility theorem \citep{Arrow1950}.
This result might be perceived as disappointing.  Depending on the problem at hand, one either has to relax marginal 
stability or to restrict the modeling to the proportional covariance model and kriging of the mean on raw data. This is a restrictive model and chances are that the covariance structure inferred from the data does not follow this form. The scientist is in an ``either/or'' situation. On the one hand, imposing C2 when a more complex model is true corresponds to forcing an inefficient estimate of the mean. On the other hand, using a more complex model of covariance than the proportional model implies that marginal stability is lost. This might lead to counter-intuitive behaviors, similar to those shown in the Introduction section. The agreement or disagreement of conclusions derived from axioms with empirical results can provide a first preliminary assessment of the appropriateness of the axioms. In this respect,  when not imposing marginal stability, the two examples shown in the Introduction can lead to results in contradiction with common sense and, perhaps, with the ultimate use of the estimates.

\medskip

This work was intended to provide new mathematical results, and to re-open  research directions. Our work is based on an axiomatic approach. Obviously, other approaches are possible. This work must be considered as complementary to the usual model-based approach, based 
on log-ratios. Instead of starting from a model and exploring its properties, we start from a set of properties (the axioms) 
and derive the class of models satisfying this set of axioms. 
Axioms are sets of properties we choose, and this choice can be discussed.
In Egozcue and Pawlowski-Glahn (2011), a principle of coherence is chosen, from which the Aitchison geometry follows. 
In essence, this is also an axiomatic approach.
Is the ``principle of coherence''  a better axiom than marginal stability? This is an interesting 
debate and an open question that would necessitate to first provide an operative definition of ``better''. 

Recent work \citep[p.167]{ScealyWelsh2014} has shown that ``subcompositional coherence [...] is essentially arbitrary and assertions on which it is based are much less natural than has been claimed. [...]''. Moreover, ``the subcompositional coherence principle actually excludes all known methods of analysing compositional data, including the log-ratio methods it was intended to privilege.''

Log-ratio analysis remains, among others, a useful tool for the statistical analysis of compositional data for some kind of statistical problems. 
Usual geostatistical tools such as covariances, variograms and kriging can be redefined in Hilbert spaces defined in the so-called Aitchison simplicial geometry \citep{Pawlowsky2002,Tolosana2011}.
An interesting development would to reformulate an axiomatic approach in the 
simplex $\BS^{p-1}$ equipped with the isometric log-ratio transform \citep{Egozcue2003} and 
check whether a result similar to Theorem \ref{theo:maintheo}
holds.

\section*{Acknowledgements} We are truly indebted to one Advisory Editor and to the Editor-in-Chief for their very constructive comments, which helped to improve this paper.

\bibliographystyle{amsalpha}

\newpage

\subsection*{Appendix A: Proof of Theorem \ref{theo:maintheo}}

Our definition of the simplex is a closed set,
i.e., some compositions are allowed to be equal to 0. 
While this might be a problem for the definition of log-ratios, it will not
be a problem for us.  Clearly the conditions are necessary: it is simple to check that \eqref{eq:Mwav} 
satisfies conditions C1-C3 for part A) and C1-C2 for part B).

\medskip

We now prove sufficiency.
By marginal stability, we have
$$
M^{k}(\bx_{1}, \ldots, \bx_{n})  =  F_{k}(x_{1}^{k}, \ldots, x_{n}^{k}) 
$$
for some function $F_{k}: \ [0,1]^{n} \rightarrow \BR$, for any $k \in \{ 1, \ldots, p \}$. If we assume continuity, as in part A), then, by the Extreme Value Theorem, $F_{k}$ is bounded since $x_{i}^k \in [0,1]$ for all $i=1,\dots,n$. If we assume $\bM(\bx_1,\dots,\bx_n) \in \BS^{p-1}$, as in part B), then $F_{k}$ is also bounded.
Choose any  $l,l',l'' \in \{1, \ldots, p\}$ 
and suppose $x_{i}^{k}=0$ for all $k \neq l,l',l''$ and all $i \in \{1, \ldots, n\}$.
Since $\sum_{k=1}^{p} M^{k}(\bx_{1}, \ldots, \bx_{n}) =1$ and  
$x_{i}^{l''} = 1 - x_{i}^{l} - x_{i}^{l'}$ for all $i \in \{1, \ldots, n\}$, it is the case that
\begin{equation}
\label{bigeq}
F_{l}(x_{1}^{l}, \ldots, x_{n}^{l}) + F_{l'}(x_{1}^{l'}, \ldots, x_{n}^{l'}) 
+ F_{l''}(1 - x_{1}^{l} - x_{1}^{l'}, \ldots, 1 - x_{n}^{l} - x_{n}^{l'}) =1.
\end{equation}
Let us define the mapping $G: \ [0,1]^{n} \rightarrow \BR$ by $G(u_{1}, \ldots, u_{n}) = 1-F_{l''}(1-u_{1}, \ldots, 1-u_{n})$.
Equation \eqref{bigeq} then becomes 
\begin{equation}
\label{g+h=d}
F_{l}(x_{1}^{l}, \ldots, x_{n}^{l}) + F_{l'}(x_{1}^{l'}, \ldots, x_{n}^{l'}) 
= G( x_{1}^{l} + x_{1}^{l'}, \ldots,  x_{n}^{l} + x_{n}^{l'}),
\end{equation}
for all $x_{1}^{l} , x_{1}^{l'} \in [0,1]$ such that $x_{1}^{l} + x_{1}^{l'} \leq 1$. 
In particular, it holds for all $x_{1}^{l} , x_{1}^{l'} \in [0,1/2]$. 
Equation~\eqref{g+h=d} is a generalized Pexider equation and, because $F_l$, $F_{l'}$ and $G$ are bounded, its unique 
solution is
\begin{align*}
F_{l}(u_{1}, \ldots, u_{n}) &=  \lambda_{1} u_{1} + \ldots + \lambda_{n} u_{n} + \gamma_{l},& \ & \forall u_{1}, \ldots, u_{n} \in [0,1/2],\\
F_{l'}(u_{1}, \ldots, u_{n}) &=  \lambda_{1} u_{1} + \ldots + \lambda_{n} u_{n} + \gamma_{l'} ,& \ & \forall u_{1}, \ldots, u_{n} \in [0,1/2],\\
G( u_{1}, \ldots,  u_{n} ) &=  \lambda_{1} u_{1} + \ldots + \lambda_{n} u_{n} + \gamma_{l} + \gamma_{l'},& \ & \forall u_{1}, \ldots, u_{n} \in [0,1],
\end{align*}
for some real numbers $\lambda_{1}, \ldots, \lambda_{n}, \gamma_{l}$ and $\gamma_{l'}$ \cite[p.302]{Aczel1966}. The expression for $G$ yields $F_{l''}( u_{1}, \ldots,  u_{n} ) = 1- G(1-u_{1}, \ldots, 1-  u_{n}) = \lambda_{1} u_{1} + \ldots + \lambda_{n} u_{n} + \beta_{l''}$ for all $u_{1}, \ldots, u_{n} \in [0,1]$ and for some real $\beta_{l''}$.

Since our choice of components $l,l'$ and $l''$ in the above reasoning is arbitrary, we obtain 
$M^{k}(\bx_{1}, \ldots, \bx_{n})  =F_{k}(x_{1}^{k}, \ldots, x_{n}^{k}) =  \sum_{i=1}^{n} \lambda_{i} x_{i}^{k} + \beta_{k}$, for all $k = 1, \ldots, p$ and all $x_{1}^{k}, \ldots, x_{n}^{k}$ in $[0,1]$. By reflexivity, $F_{k}(u, \ldots, u)  =  \sum_{i=1}^{n} \lambda_{i} u + \beta_{k} = u$, for all $u \in [0,1]$ and for all $k = 1, \ldots, p$. This is possible only if $\beta_{k}=0$ for all $k = 1, \ldots, p$ and $\sum_{i=1}^{n} \lambda_{i} = 1$. 
\epr

\newpage

\subsection*{Appendix B: Proof of Theorem \ref{theo:maintheo2}}

In the following, $\Un_{n}$ denotes a vector of ones of length $n$, $\bI_n$ denotes the identity matrix of dimension $n$ and 
$\Zero_{p,q}$ denotes a $p \times q$ matrix of zeros. If $\bA$ is a $m \times n$ matrix and $\bB$ is a $p \times q$ matrix, the
Kronecker product $\bA \otimes \bB$ is the $mp \times nq$ block matrix
$$\bA \otimes \bB = \left( \begin{array}{ccc}  a_{11} \bB & \cdots & a_{1n} \bB \\ 
\vdots & \ddots & \vdots \\
a_{m1} \bB & \cdots & a_{mn} \bB 
\end{array}  \right).$$
In particular, the  matrix $\bJ = \bI_p \otimes \Un_n$ is the $np \times p$ matrix
$$ \left( \begin{array}{ccc} \Un_{n} & \cdots & \Zero_{n,1} \\
\vdots & \ddots & \vdots \\
\Zero_{n,1} & \cdots &  \Un_{n}
\end{array}  \right).$$

\medskip

\begin{enumerate}
\item[A)]  We first consider unconstrained kriging. 

For each variable $k=1,\dots,p$, the kriging of the mean, $\hat{m}_k$, is a linear combination of the data
$$\hat{m}_k = \sum_{l=1}^p \bX_l^\top  \blambda^k_l,$$  
where $\bX_l = (x_{1,l},\dots,x_{n,l})^\top$ and $\blambda^k_l = (\lambda^k_{1,l},\dots,\lambda^k_{n,l})^\top$. 
Unbiasedness conditions impose
$$\Un_{n}^\top \blambda^k_k=1\ \ \hbox{and} \ \ \Un_{n}^\top \blambda^k_l=0,\ \hbox{for}\ l \neq k.$$
Let $\blambda^k$ be the stacked $np$-vector $\blambda^k = ( (\blambda^k_1)^\top,\dots,(\blambda^k_p)^\top)^\top$ and
let  $\bLambda =(\blambda^1,\dots,\blambda^p)$ be the $(np \times p)$ matrix of kriging weights.
When solved simultaneously, the $p$  kriging equations are, in matrix notation,
\begin{equation}
  \label{eq:kriging}
  \left(\begin{array}{cc} \bC & \bJ \\ \bJ^t & \Zero_{p,p} \end{array} \right) \left( \begin{array}{c}  \bLambda \\ -\bmu \end{array}  \right) 
= \left( \begin{array}{c}  \Zero_{np,p} \\ \bI_p \end{array}  \right),
\end{equation}
where  $\bmu$ is the $p \times p$ matrix of Lagrange multipliers. The matrix $\bC$ arises from a valid model of covariance functions. If one excludes multiple
values at the same location, it is invertible.
The general solution for $\bLambda$ therefore satisfies
\begin{equation}
  \label{eq:sol_kriging}
\bC \bLambda  =  \bJ \bmu, 
\end{equation}
where $\bmu=(\bJ^\top \bC^{-1} \bJ)^{-1}$. We will denote $\mu_{kl}$ its elements,
with $1 \leq k,l \leq p$.

For each $k=1,\dots,p$, we wish to impose that $\blambda^k$ is a vector of zeros except at coordinates corresponding to the 
$k$-th variable where the weights  are equal to a common vector $\blambda$, i.e.,
$\blambda^k = (\Zero_{1,n(k-1)},\blambda^\top,\dots,\Zero_{1,n(p-k)})^\top$. Hence,
$\bLambda = \bI_p \otimes \blambda$. Thus, Eq.~(\ref{eq:sol_kriging}) becomes
\begin{equation}
  \label{eq:cov_cond}
\bC (\bI_p \otimes \blambda) =  \bJ \bmu,  
\end{equation}
Then, Eq.~(\ref{eq:cov_cond}) is equivalent to
\begin{equation}
  \label{eq:cov_cond2}
  \bC_{kl} \blambda = \mu_{kl} \Un_n, \ \forall \ 1 \leq k,l \leq p.
\end{equation}
Since $\bC$ is invertible, $\bC_{kk}$ is invertible and $\mu_{kk} \neq 0$, for all $k=1,\dots,p$. Hence, plugging 
$\mu_{kk} \bC_{kk}^{-1} \Un_n = \blambda$ into Eq.~(\ref{eq:cov_cond2}) leads to
$$\frac{\mu_{kk}}{\mu_{ll}} \bC_{kk}^{-1} \bC_{ll} = \bI_n, \ \forall \ 1 \leq k,l \leq p.$$
This condition shows that $\bC_{kk} = a_{kk} \bR$, where $\bR$ is a correlation matrix with $a_{kk}>0$. 
With a similar argument, one can show that $\bC_{kl} = a_{kl} \bR$ when $k \neq l$.
In conclusion, there is a single correlation matrix for describing all direct and cross covariance matrices, 
$$\bC = \bSigma \otimes \bR.$$
In other words, the model is proportional.

\item[B)] We now impose non-negativity of the kriging weights. This constrained kriging is the solution of the quadratic system
  \begin{eqnarray*}
    & \min_{\blambda} &  \sum_{k=1}^p \blambda^\top \bC_{kk} \blambda\\
 & \hbox{\rm s.t.} & \blambda^\top  \Un_n = 1 \\
& & \blambda \geq  \Zero_n,
  \end{eqnarray*}
where the last inequality must be satisfied componentwise. The Kuhn-Tucker stationary conditions \citep{Griva2008} corresponding to this system are 
\begin{eqnarray*}
  \bC_{kk} \blambda - \mu_k \Un_n - \balpha & = 0 & k=1,\dots,p\\
\blambda^\top  \Un_n  & = & 1\\
\alpha_i & \geq 0 & i=1,\dots,n\\
\alpha_i \lambda_i & = 0 &  i=1,\dots,n,
\end{eqnarray*}
where $\balpha = (\alpha_1,\dots,\alpha_n)^\top$ and $\bmu = (\mu_1,\dots,\mu_p)^\top$ are the $n+p$ Lagrange multipliers. 
A non-negativity constraint is said to be active if $\alpha_i>0$ and non active when $\alpha_i=0$.

The Lagrange multipliers $\balpha$, and hence the set of active constraints, are identical for all $k=1,\dots,p$. Let us re-order the dataset such that the first $m$ elements, with $1 < m \leq n$ correspond to inactive constraints, i.e. $\alpha_i=0$ and $\lambda_i > 0$ for $i=1,\dots,m$. 
Let us denote $\blambda_m$ the corresponding vector of non-null kriging weights and $\bC_{kk}^{m,m}$ the corresponding $m\times m$ matrix with the $m$ first rows and columns of $\bC_{kk}$. Let us also denote $\bC_{kk}^{m,n-m}$ the 
 $m\times (n-m)$ matrix with elements from the $m$ first rows and $(n-m)$ last columns. Then, 
 $\blambda_m$ is solution of the kriging system
\begin{eqnarray}
  \bC_{kk}^{m,m} \blambda_m - \mu_k \Un_m & = 0 & k=1,\dots,p \label{eq:KT1}\\
\blambda_m^\top  \Un_m  & = & 1 \label{eq:KT2}\\
 \blambda_m \bC_{kk}^{m,n-m}  - \mu_k \Un_{n-m} - \balpha_{n-m} & = 0 & k=1,\dots,p \nonumber \\ 
\alpha_i & \geq 0 & i=m+1,\dots,n. \nonumber 
\end{eqnarray}
Following arguments similar to those in part A), Eqs. (\ref{eq:KT1}) and (\ref{eq:KT2}) induce that all covariance matrices  $\bC_{kk}^{m,m}$ must be
proportional to each other, i.e. $\bC_{kk}^{m,m} = a_{kk} \bR^{m,m}$. This condition is  satisfied for all $m$ if and only if $\bC_{kk} = a_{kk} \bR$,
i.e.  if and only if $\bC = \bSigma \otimes \bR.$
\end{enumerate}
 \epr

\end{document}